\documentclass{article}
\usepackage{algorithm}
\usepackage{algorithmic}
\def\noheaderplainsetup{%

\topmargin=-10pt \headheight=0pt \headsep=0pt  \oddsidemargin=0pt \evensidemargin=0pt  \textheight=9.23truein \textwidth=6.5truein}

\noheaderplainsetup

\begin{document}


\newcommand{\lmt}{\ell}
\newcommand{\col}[1]{\mbox{$#1$:}}
\newcommand{\chess}{\mbox{\em Chess}}
\newcommand{\checkers}{\mbox{\em Checkers}}
\newcommand{\elzz}[2]{ \langle #1\rangle\hspace{-2pt} \downarrow\hspace{-2pt} #2}
\newcommand{\elz}[1]{\mbox{$\parallel\hspace{-3pt} #1 \hspace{-3pt}\parallel$}}
\newcommand{\qelz}[1]{\mbox{$| #1 |$}}
\newcommand{\emptyrun}{\langle\rangle} 
\newcommand{\oo}{\bot}            
\newcommand{\pp}{\top}            
\newcommand{\xx}{\wp}               
\newcommand{\win}[2]{\mbox{\bf Wn}^{#1}_{#2}} 
\newcommand{\seq}[1]{\langle #1 \rangle}           


\newcommand{\plus}{\mbox{\hspace{1pt}\raisebox{0.05cm}{\tiny\boldmath $+$}\hspace{1pt}}}
\newcommand{\mult}{\mbox{\hspace{1pt}\raisebox{0.05cm}{\tiny\boldmath $\times$}\hspace{1pt}}}
\newcommand{\mminus}{\mbox{\hspace{1pt}\raisebox{0.05cm}{\tiny\boldmath $-$}\hspace{1pt}}}
\newcommand{\equals}{\mbox{\hspace{1pt}\raisebox{0.05cm}{\tiny\boldmath $=$}\hspace{1pt}}}
\newcommand{\notequals}{\mbox{\hspace{1pt}\raisebox{0.05cm}{\tiny\boldmath $\not=$}\hspace{1pt}}}

\newcommand{\mless}{\mbox{\hspace{1pt}\raisebox{0.05cm}{\tiny\boldmath $<$}\hspace{1pt}}}
\newcommand{\mgreater}{\mbox{\hspace{1pt}\raisebox{0.05cm}{\tiny\boldmath $>$}\hspace{1pt}}}
\newcommand{\mleq}{\mbox{\hspace{1pt}\raisebox{0.05cm}{\tiny\boldmath $\leq$}\hspace{1pt}}}
\newcommand{\mgeq}{\mbox{\hspace{1pt}\raisebox{0.05cm}{\tiny\boldmath $\geq$}\hspace{1pt}}}

\newcommand{\clt}{\mbox{\bf CL13}}
\newcommand{\clte}{\mbox{\bf CL14}}
\newcommand{\cltee}{\overline{\clte}}
\newcommand{\intimpl}{\mbox{\hspace{2pt}$\circ$\hspace{-0.14cm} \raisebox{-0.043cm}{\Large --}\hspace{2pt}}}
\newcommand{\sintimpl}{\mbox{\hspace{2pt}\raisebox{0.033cm}{\tiny $ | \hspace{-4pt} >$}\hspace{-0.14cm} \raisebox{-0.039cm}{\large --}\hspace{2pt}}}
\newcommand{\ade}{\mbox{\Large $\sqcup$}\hspace{1pt}}      
\newcommand{\ada}{\mbox{\Large $\sqcap$}\hspace{1pt}}      
\newcommand{\sst}{\mbox{\raisebox{-0.07cm}{\scriptsize $-$}\hspace{-0.2cm}$\pst$}}
\newcommand{\scost}{\mbox{\raisebox{0.20cm}{\scriptsize $-$}\hspace{-0.2cm}$\pcost$}}
\newcommand{\sqc}{\mbox{\hspace{2pt}\small \raisebox{0.0cm}{$\bigtriangleup$}\hspace{2pt}}}
\newcommand{\sqci}{\mbox{\scriptsize \raisebox{0.0cm}{$\bigtriangleup$}}}
\newcommand{\sqd}{\mbox{\hspace{2pt}\small \raisebox{0.06cm}{$\bigtriangledown$}\hspace{2pt}}}
\newcommand{\sqdi}{\mbox{\scriptsize \raisebox{0.05cm}{$\bigtriangledown$}}}
\newcommand{\sqe}{\mbox{\large \raisebox{0.07cm}{$\bigtriangledown$}}}
\newcommand{\sqa}{\mbox{\large \raisebox{0.0cm}{$\bigtriangleup$}}}
\newcommand{\mld}{\vee}    
\newcommand{\mlc}{\wedge}  
\newcommand{\tgd}{\mbox{\hspace{2pt}$\vee$\hspace{-1.29mm}\raisebox{0.1mm}{\rule{0.13mm}{2mm}}\hspace{5pt}}}    
\newcommand{\tgc}{\mbox{\hspace{2pt}$\wedge$\hspace{-1.29mm}\raisebox{0.02mm}{\rule{0.13mm}{2mm}}\hspace{5pt}}}    
\newcommand{\tge}{\hspace{1pt}\mbox{\Large $\vee$\hspace{-1.84mm}\raisebox{0.1mm}{\rule{0.13mm}{3.0mm}}\hspace{6pt}}}
\newcommand{\tga}{\mbox{\hspace{1pt}\Large $\wedge$\hspace{-1.84mm}\raisebox{0.02mm}{\rule{0.13mm}{3.0mm}}\hspace{6pt}}}
\newcommand{\tgpst}{\mbox{\raisebox{-0.01cm}{\scriptsize $\wedge$}\hspace{-4pt}\raisebox{0.06cm}{\small $\mid$}\hspace{2pt}}}
\newcommand{\tgpcost}{\mbox{\raisebox{0.12cm}{\scriptsize $\vee$}\hspace{-3.8pt}\raisebox{0.04cm}{\small $\mid$}\hspace{2pt}}}
\newcommand{\tgst}{\mbox{\raisebox{-0.05cm}{$\circ$}\hspace{-0.12cm}\raisebox{0.05cm}{\small $\mid$}\hspace{2pt}}}
\newcommand{\tgcost}{\mbox{\raisebox{0.12cm}{$\circ$}\hspace{-0.12cm}\raisebox{0.04cm}{\small $\mid$}\hspace{2pt}}}
\newcommand{\tgpi}{\mbox{\hspace{2pt}\raisebox{0.033cm}{\tiny $>$}\hspace{-0.28cm} \raisebox{-2.3pt}{\LARGE --}\hspace{2pt}}}
\newcommand{\tgbi}{\mbox{\hspace{2pt}$\circ$\hspace{-0.26cm} \raisebox{-2.3pt}{\LARGE --}\hspace{2pt}}}
\newcommand{\mle}{\mbox{\hspace{1pt}\Large $\vee$}\hspace{1pt}}    
\newcommand{\mla}{\mbox{\hspace{1pt}\Large $\wedge$}\hspace{1pt}}  
\newcommand{\add}{\hspace{0pt}\sqcup}                      
\newcommand{\adc}{\hspace{0pt}\sqcap}                      
\newcommand{\gneg}{\neg}                  
\newcommand{\rneg}{\neg}               
\newcommand{\pneg}{\neg}               
\newcommand{\mli}{\rightarrow}                     
\newcommand{\intf}{\$}               
\newcommand{\tlg}{\bot}               
\newcommand{\twg}{\top}               

\newcommand{\pst}{\mbox{\raisebox{-0.01cm}{\scriptsize $\wedge$}\hspace{-4pt}\raisebox{0.16cm}{\tiny $\mid$}\hspace{2pt}}}
\newcommand{\cla}{\mbox{\large $\forall$}\hspace{1pt}}      
\newcommand{\cle}{\mbox{\large $\exists$}\hspace{1pt}}        
\newcommand{\pintimpl}{\mbox{\hspace{2pt}\raisebox{0.033cm}{\tiny $>$}\hspace{-0.18cm} \raisebox{-0.043cm}{\large --}\hspace{2pt}}}
\newcommand{\pcost}{\mbox{\raisebox{0.12cm}{\scriptsize $\vee$}\hspace{-4pt}\raisebox{0.02cm}{\tiny $\mid$}\hspace{2pt}}}
\newcommand{\st}{\mbox{\raisebox{-0.05cm}{$\circ$}\hspace{-0.13cm}\raisebox{0.16cm}{\tiny $\mid$}\hspace{2pt}}}
\newcommand{\cost}{\mbox{\raisebox{0.12cm}{$\circ$}\hspace{-0.13cm}\raisebox{0.02cm}{\tiny $\mid$}\hspace{2pt}}}


\newtheorem{theoremm}{Theorem}[section]
\newtheorem{conjecturee}[theoremm]{Conjecture}
\newtheorem{exercisee}[theoremm]{Exercise}
\newtheorem{definitionn}[theoremm]{Definition}
\newtheorem{lemmaa}[theoremm]{Lemma}
\newtheorem{propositionn}[theoremm]{Proposition}
\newtheorem{conventionn}[theoremm]{Convention}
\newtheorem{examplee}[theoremm]{Example}
\newtheorem{remarkk}[theoremm]{Remark}
\newtheorem{factt}[theoremm]{Fact}
\newtheorem{claimm}[theoremm]{Claim}

\newenvironment{conjecture}{\begin{conjecturee}}{\end{conjecturee}}
\newenvironment{definition}{\begin{definitionn} \em}{ \end{definitionn}}
\newenvironment{theorem}{\begin{theoremm}}{\end{theoremm}}
\newenvironment{lemma}{\begin{lemmaa}}{\end{lemmaa}}
\newenvironment{proposition}{\begin{propositionn} }{\end{propositionn}}
\newenvironment{convention}{\begin{conventionn} \em}{\end{conventionn}}
\newenvironment{remark}{\begin{remarkk} \em}{\end{remarkk}}
\newenvironment{proof}{ {\bf Proof.} }{\  \rule{2mm}{2mm} \vspace{.15in} }
\newenvironment{example}{\begin{examplee} \em}{\end{examplee}}
\newenvironment{exercise}{\begin{exercisee} \em}{\end{exercisee}}
\newenvironment{fact}{\begin{factt} \em}{\end{factt}}
\newenvironment{claim}{\begin{claimm} \em}{\end{claimm}}

\title{Implementing program extraction from CL1-proofs}
\author{Meixia Qu$^{1,2}$ \space Ke Chen$^2$\space Daming Zhu$^1$\space Junfeng Luan$^1$  \\ \\
{\small $^1$School of Computer Science and Technology, Shandong University;}\\
{\small $^2$School of Mechanical, Electrical\&Information Engineering, Shandong University at Weihai}}
\date{}
\maketitle

\begin{abstract}
{\em Computability logic} (CoL) is a formal theory of interactive computation. It understands computational problems as games played by two players: a machine and its environment, uses logical formalism to describe valid principles of computability and formulas to represent computational problems. Logic CL1 is a deductive system for a fragment of CoL. The logical vocabulary contains all of the operators of classical logic and choice operators, the atoms represent elementary games i.e. predicates of classical logic. In this paper, we present a program that takes a CL1-proof of an arbitrary formula $F$, and extract a winning strategy for $F$ from that proof then play $F$ using that strategy. We hope this paper would provide a starting point for further work in program extraction of the CoL-based arithmetic and other CoL-based applied systems.

\end{abstract}



\section{Introduction}\label{sintr}

{\em Computability logic} (CoL), introduced by Japaridze in \cite{Jap0} and extensively studied in recent years (\cite{Japi}-\cite{Japxure2} and many more), is a systematic and still-evolving formal theory of computability. In it, computational problems are seen as games between two players: a machine and its environment. Machine represents a mechanical device with fully determined behavior, while environment represents a capricious user with arbitrary behavior. Logical operators stand for operations on games, and ``truth'' is seen as existence of an algorithmic solution, i.e. of a machine's winning strategy. While winnability is a property of games, the validity is a property of a logical formula which represents an ``always computable'' problem.

The main technical goal of CoL at the current stage is to axiomatize the set of valid principles of computability or various natural fragment of that set. In recent years, there has been a rapid progress (\cite{JapCL1}-\cite{Japxure2}) in this area. Japaridze has presented the system from CL1 (\cite{JapCL1}) to CL15 (\cite{Japtam1}) so far. The language of logic CL1, which the current paper is exclusively devoted to, is a basic deductive propositional system of CoL obtained  by adding to the language of classical propositional logic two additional choice operators: disjunction ($\add$) and conjunction ($\adc$) operators, the atoms of CL1 represent elementary games i.e. predicates of classical logic. The purpose of logic is providing a tool in real life. So, it is necessary to start investigating program extraction from the various CoL-systems. In this paper, we present a program that takes a CL1-proof of an arbitrary formula $F$, extract a winning strategy for $F$ from that proof, and then play $F$ using that strategy. Moreover, we define the step of the CL1-proofs stored in a file as an input for the program and output an interactive interface during playing $F$. We hope this paper would provide a starting point for further work in program extraction from the CoL-based arithmetic and other CoL-based applied systems.

We will not reintroduce the main concepts of Computability logic in the present paper, all of them can be found in \cite{Japfi}. However, it would help the readers of this paper in understanding the underlying meanings and motivations of computability logic if the reader is familiar with that paper. In order to make this paper mathematically self-contained, we will reintroduce the relevant concepts of CL1 in our paper.

The rest of this paper is organized as follows. Some basic notation and terminology of CL1 from the earlier literature in \cite{JapCL1} will be introduced in Section \ref{s2tb}. Section \ref{s22tog} introduces an algorithm that solves the problem represented by a formula $F$ from a CL1-proof. We present the correctness of our algorithm in Section \ref{s3-3tb} and provide the complexity analysis in Section \ref{s4-3tb}. The final Section \ref{s5thr} contains some conclusive remarks.

\section{Preliminaries}\label{s2tb}
Since the rest of this paper is devoted to program extraction from CL1-proofs, we will reintroduce some necessary preliminaries in this section.
All of them also can be found in \cite{JapCL1}.
\subsection{Basic notation and terminology for CL1}\label{s2-1tb}
The language of logic CL1 is obtained by adding two extra operators $\add$ and $\adc$ in the language of classical propositional logic. It is built from the following elements:
\begin{enumerate}

\item Two {\bf logical atoms}: $\oo$, $\pp$.

\item Infinitely many {\bf non-logical atoms}: $p$, $q$, $r$, $s$, $p_1$, $p_2$, $p_3$, $...$.

\item {\bf Operators}: $\neg$, $\mli$, $\mlc$, $\mld$, $\adc$, $\add$.

\item {\bf Parentheses}: (, ).
\end{enumerate}

The {\bf Formulas} of CL1 are defined inductively by:
\begin{itemize}

\item Atoms (logical or nonlogical) are formulas.

\item If $F$ is a formula, then so is $\neg(F)$.

\item If $F_1$, $...$, $F_n$ ($n\mgeq 2$) are formulas, then so are $(F_1)\mli(F_2)$, $(F_1)\mld ... \mld(F_n)$,  $(F_1)\mlc ... \mlc(F_n)$, $(F_1)\add ... \add(F_n)$, $(F_1)\adc ... \adc(F_n)$.
\end{itemize}

Understanding $E\mli F$ as an abbreviation of $\neg E \mld F$, a {\bf positive} occurrence of a subformula is one that is in the scope of an even number of $\neg$'s. Otherwise, the occurrence is {\bf negative}.

A {\bf surface occurrence} of a subformula means an occurrence that is not in the scope of a choice ($\add$ or $\adc$) operator.

A CL1-formula is said to be {\bf elementary} iff it does not contain the choice operators.

The {\bf elementarization} of a CL1-formula is the result of replacing, in it, every surface occurrence of the form $F_1\add ... \add F_n$ by $\oo$ , and every surface occurrence of the form $F_1\adc ... \adc F_n$ by $\pp$.

A CL1-formula is said to be {\bf stable} iff its elementarization is valid in classical logic, otherwise the formula is {\bf instable}.

$F${\bf -specification} of $O$, where  $F$ is a formula and $O$ is a surface occurrence in $F$, is a string $\alpha$ which can be defined by:
\begin{itemize}

\item $F$-specification of the occurrence in itself is the empty string.

\item If $F$ = $\neg G$, then $F$-specification of an occurrence that happens to be in $G$ is the same as the $G$-specification of that occurrence.

\item If $F$ is $G_1\mlc ... \mlc G_n$, $G_1\mld ... \mld G_n$, or $G_1\mli G_2$, then $F$-specification of an occurrence that happens to be in $G_i$ is the string $i.\alpha$, where $\alpha$ is the $G_i$-specification of that occurrence.

\end{itemize}
\subsection{The rules of CL1}\label{s2-2tb}
CL1 has the following two rules, with $E$, $F$ standing for CL1-formulas and $\vec H$ for a set of CL1-formulas:

Rule (a): ${\vec H}\vdash F$, where $F$ is stable and, whenever $F$ has a positive (resp. negative) surface occurrence of $G_1\adc ... \adc G_n$ (resp. $G_1\add ... \add G_n$), for each i$\in \{1,...,n\}$, $\vec H$ contains the result of replacing in $F$ that occurrence by $G_i$.

Rule (b): $E\vdash F$, where $E$ is the result of replacing in $F$ a negative (resp. positive) surface occurrence of $G_1\adc ... \adc G_n$ (resp. $G_1\add ... \add G_n$) by $G_i$ for some i$\in \{1,...,n\}$.

In this paper, the following algorithm will read the CL1-proofs as its input. We assume that the proof consists of a sequence split by the line break without other redundancies. Below we define the formal format of every proof step in the sequence. Meanwhile, we refer to formulas in the proofs as ``proof formulas''.

\begin{definition}\label{defc21}
The {\bf step} is defined as follows. There are three cases to be considered.

{\em Case 1}: There is no premise when using rule (a). In other words, this proof formula is an axiom. The step should be:

linenumber. proof formula, rule a, no premise

{\em Case 2}: There are two premises when using rule (a). The step should be:

linenumber. proof formula, rule a, linenumber of premise1  linenumber of premise2

{\em Case 3}: If a proof formula can be derived by rule (b), The step should be:

linenumber. proof formula, rule b, linenumber of premise

\end{definition}

\begin{examplee}\label{ex01}

$CL1 \vdash ((p\adc q)\mlc(p\adc q))\mli(p\adc q)$

$p$, $q$ represent distinct non-logical atoms.
\end{examplee}
\begin{enumerate}

\item $(p\mlc p)\mli p$, rule a, no premise

\item $(q\mlc q)\mli q$, rule a, no premise

\item $((q\adc p)\mlc p)\mli p$, rule b, 1

\item $((p\adc q)\mlc (q\adc p))\mli p$, rule b, 3

\item $((p\adc q)\mlc q)\mli q$, rule b, 2

\item $((p\adc q)\mlc (p\adc q))\mli q$, rule b, 5

\item $((p\adc q)\mlc (p\adc q))\mli (p\adc q)$, rule a, 4 6

\end{enumerate}

\begin{examplee}\label{ex02}
$CL1 \vdash p\mli (r\adc q)$

$p$, $q$, $r$ represent distinct non-logical atoms.
\end{examplee}
\begin{enumerate}

\item $p\mli q$, rule a, no premise

\item $p\mli r$, rule a, no premise

\item $p\mli (r\adc q)$, rule a, 1 2

\end{enumerate}

\section{An algorithm for solving the problem represented by a CL1 formula}\label{s22tog}

In this section we will introduce an algorithm that solves the problem represented by a formula $F$ from a CL1-proof of $F$. The algorithm contains two stages. First stage is to select the desirable data structures and parse the CL1-proofs. We will finally construct a syntax analysis tree and store the information about the justification of every step in CL1-proofs. This storage is necessary for us to extract a winning strategy during the interactive computation. All the relative data structures and operations in the construction of a tree will be introduced in section \ref{s3-1tb}. The second stage is to present the elimination operation. We will introduce an algorithm to eliminate the choice operators in a formula and provide a true interactive computation between machine and environment in section \ref{s3-2tb}.

\subsection{Data structures and operations on the CL1-proofs}\label{s3-1tb}
In the first stage of the algorithm, we will construct a syntax analysis tree to represent a formula. If the node is a leaf, then the value of this node is the non-logical atoms. Otherwise, the value of a node will be either operators or parenthesis. There are three cases to be considered depending on the proof formula derived by different rules during the construction of a tree.

{\em Case 1}: A proof formula is derived by rule (a) with no premise. In this case, we read the proof formula, construct a syntax tree directly and return this tree.

{\em Case 2}: A proof formula is derived by rule (a) with premises existing in the previous step. In this case, there will be an extra operation before the construction. Assume $\alpha$ = $\beta i$, where $\beta$ is a $E$-specification of a positive (resp. negative) surface occurrence of a subformula $G_1\adc ... \adc G_n$ (resp. $G_1\add ... \add G_n$) and i$\in \{1,...,n\}$. Let $H$ be the result of substituting the above occurrence by $G_i$ in $E$. If $H$ is isomorphic with one of the premises of $E$, we store $\alpha$ and the corresponding premise $H$ as one valid choice for environment during the interactive computation. This operation will guarantee that we can find all valid choices for the environment.

{\em Case 3}: A proof formula is derived by rule (b) with a premise existing in the previous step. In this case, we will do the same extra operation for the syntax analysis tree. Assume $\alpha$ = $\beta i$, where $\beta$ is a $E$-specification of a negative (resp. positive) surface occurrence of a subformula $G_1\adc ... \adc G_n$ (resp. $G_1\add ... \add G_n$) by $G_i$ for some i$\in \{1,...,n\}$. Let $H$ be the result of substituting the above occurrence by $G_i$ in $E$. If $H$ is isomorphic with the premise of $E$. We store $\alpha$ and its corresponding premise $H$ as one valid choice for machine during the interactive computation. This operation will guarantee that we can find all valid choices for the machine.

\subsection{Elimination operation}\label{s3-2tb}

After the construction of a tree, we will describe an elimination operation as the second stage. The goal of this operation is to eliminate all choice operators occurring in the tree. Meanwhile, we also can output an interactive user interface by traversing the tree.

\begin{algorithm}
\caption{Elimination Operation}
\label{121}
\begin{algorithmic}[1]
\REQUIRE~~\\ A CL1 formula $F$ and $F$'s syntax tree $T$
\ENSURE~~\\Interactive computation of a CL1 formula
   \STATE current formula $E$ $\leftarrow$ $F$
   \STATE traverse $T$ in order
   \WHILE{$E$ is a proof formula}

        \IF{$E$ is derived by Rule (a)}
                 \STATE {Waiting until the environment make a move $\beta$}
                 \FOR {each valid choice for the environment $\alpha$}
                     \IF {$\alpha$ is equal to $\beta$}

                         \STATE {$E$ $\leftarrow$ $E$'s premise which is specified by $\beta$ in $E$}
                         \STATE {break}
                     \ENDIF

                 \ENDFOR
     \ELSIF{$E$ is derived by Rule (b)}

        \STATE {$\beta \leftarrow$ $E$'s an arbitrary valid choice for the machine}

        \STATE {$E$ $\leftarrow$ $E$'s premise which is specified by $\beta$ in $E$}


     \ENDIF

     \STATE traverse $T$ in order

   \ENDWHILE

\end{algorithmic}
\end{algorithm}
\section{Correctness of the algorithm}\label{s3-3tb}
Let us verify the correctness of our algorithm. The main operation in the first stage of our algorithm is to judge the isomorphism between two trees. Because \cite{JapWen} has verified the correctness of tree isomorphism, we only present the correctness of Algorithm \ref{121}. We use the loop invariant. At the start of each iteration in {\bf while} loop of lines 4-16 in Algorithm \ref{121}, $E$ is always a CL1 formula.

Initialization: $E \leftarrow F$ and $F$ must be a CL1 formula, so the invariant is true.

Maintenance: $E$ will be substituted by the $E$'s premise after a choice which was made by the environment in line 8 or the machine in line 14 respectively. From the rules of CL1 we know that each premise of $E$ is also a CL1 formula. So the loop invariant can be maintained at all times thereafter.

Termination: Because Algorithm \ref{121} will eliminate one choice operator in each iteration and there are definitely many choice operators in $E$, the loop will be terminated until there are no choice operators in $E$. Finally, $E$ will be an elementary formula.

\section{Computational complexity analysis}\label{s4-3tb}

We will present an analysis of our algorithm on time complexity which described in section \ref{s22tog}. From the BULT algorithm in \cite{JapWen}, we can implement the isomorphism testing in $O(n)$. There is some extra analysis for the second stage because we should additionally consider the interactive complexity during the interactive computation. We call the time complexity as the classical time complexity for an algorithm without considering the interactive complexity. Algorithm \ref{121} will iterate $O(n)$ times in {\bf while} loop in order to eliminate the choice operators for $F$. If a proof formula $E$ is derived by Rule (a), the algorithm will iterate all the valid choices for the environment, the cost of the total loop is $O(n)$. If $E$ is derived by Rule (b), the algorithm will make an arbitrary valid choice for the machine, the cost on lines 13-14 is $O(1)$. After that, algorithm will traverse the current syntax tree to output the formula, it will take time O(n). So the cost of the loop body on lines 4-16 is O(n). Therefore, the classical time complexity of Algorithm \ref{121} is $O(n^2)$. From the definition 5.2 of \cite{JapCL12}, we know that the interactive complexity in Algorithm \ref{121} is constant time. So the overall time complexity is $O(n^2)$, and we can implement this algorithm in a polynomial time.


\section{Conclusion} \label{s5thr}
The work in CoL in recent years has been mainly focused on purely theoretical aspects. In this paper, we propose a program extraction from CL1-proofs which can be implemented in polynomial time. This work provides us a new research domain to search the algorithm and implement it as an interactive computation in CoL. We hope this paper well provide a starting point for further work in program extraction from the CoL-based arithmetic and other CoL-based applied systems.


\end{document}